# Complexity Synchronization

Korosh Mahmoodi[1,2*], Scott E. Kerick[1], Paolo Grigolini[2], Piotr J. Franaszczuk[1,3] and Bruce J. West[2]

1. US Combat Capabilities Command, Army Research Laboratory, Aberdeen Proving Ground, 21005, MD, USA.
2. Center for Nonlinear Science, University of North Texas, Denton, 76203, TX, USA.
3. Department of Neurology, Johns Hopkins University School of Medicine, Baltimore, 21287, MD, USA.

* Corresponding author. E-mail: koroshmahmoodi@my.unt.edu
  Contributing authors: scott.e.kerick.civ@army.mil; paolo.Grigolini@unt.edu; piotr.j.franaszczuk.civ@army.mil; brucejwest213@gmail.com

*Since the turn of the century Network Science and Complexity Theory have been growing dramatically and their nexus has led to profoundly different ways of thinking about physiology, health, disease, and medicine in general from the modeling based on the Newtonian paradigm. The observational ubiquity of inverse power law spectra (IPL) in complex phenomena entails theory for dynamic fractal phenomena capturing their fractal dimension, dynamics, and statistics[1]. These and other properties are consequences of the complexity resulting from nonlinear dynamic networks collectively summarized for biomedical phenomena as the Network Effect (NE)[2] or focused more narrowly as Network Physiology[3]. The NE is often described by homogeneous scaling variables with power law scaling having index δ determined by the fractal dimension of the time series[4] being a direct measure of the network's complexity[5]. Herein we address the measurable consequences of the NE on time series generated by different parts of the brain, heart, and lung organ networks, which are directly related to their inter-network and intra-network interactions. Moreover, these same physiologic organ networks have been shown to generate crucial event (CE) time series[6], and herein are shown, using modified diffusion entropy analysis (MDEA)[5], to have scaling indices with quasiperiodic changes in complexity, as measured by scaling indices, over time. Such time series are generated by different parts of the brain, heart, and lung organ networks, and the results do not depend on the underlying coherence properties of the associated time series[5] but demonstrate a generalized synchronization of complexity. This high-order synchrony among the scaling indices of EEG (brain), ECG (heart), and respiratory time series is governed by the quantitative interdependence of the multifractal behavior of the various physiological organs' network dynamics. This consequence of the NE opens the door for an entirely general characterization of the dynamics of complex networks in terms of complexity synchronization (CS) independently of the scientific, engineering, or technological context. CS is truly a transdisciplinary effect.*

Synchronization is today identified as the mechanism needed to coordinate activities among events in any complex, multilevel, multielement dynamic network. However, as the network becomes more complex so does the changing concept of synchronization. This is particularly true of the amazingly complex organ network structure of the human body and the need to coordinate activities across vastly different time scales, from the microscopic time scales of the neural networks within the brain, to the mesoscopic time scales of the cardiac and respiratory organ networks, to the macroscopic time of circadian rhythms.

We show that the complexity (scaling) of brainwave time series data is multifractal, as are the respiratory and cardiovascular time series, and that the multifractal scaling of the three are synchronous. The multifractal behavior of these time series has been identified using pairwise correlation of time series to identify an appropriate mechanism[6]. The change in fractal scaling of the time series indicates the changing complexity of the organ networks as various physiological functions are performed. In the language of network science, the more complex network is the driver, and the less complex network is the driven, but these roles change with the functions being performed and can change in time, as well. Information is readily transported within overlapping memory areas of the heterogeneously complex brain and at any point in time a given region of the brain (driven) can receive information from sensor networks, process that information and transmit the processed signal (driver), to an appropriate physiological organ network (driven) for action, depending on their function and instantaneous relative complexities. This hierarchy of the average complexity is subsequently revealed by the way in which the multifractal nature of each of these three interacting organ networks influence one another over time. The relative width of the multifractal spectra determines the driver (greater width) and the driven (smaller width)[2].

Recent advances in data processing techniques have revealed a new kind of synchronization, complexity synchronization (CS), based on crucial events (CEs)[5], which enables the detection of synchrony among complex dynamic organ networks operating on different time scales and not necessarily in stationary regimes, as we shall show. A sequence of CEs is a renewal statistical process generated by an inverse power law (IPL) probability density function (PDF). If the time interval between events is $\tau$ the IPL PDF is $\psi(\tau) \propto \tau^{-\mu}$ and the IPL index µ is in the domain 1<µ<3. Asymptotically, the generated process of CEs is ergodic for 2<µ<3, with a finite average time interval between CEs and is non-ergodic for 1<µ<2, with an infinite average time interval between CEs. The scaling (complexity) of healthy brainwave time series is shown to be multifractal, as are the healthy respiratory and healthy cardiovascular time series. Moreover, the multifractal scaling of these three healthy data sets are determined to be in synchrony with one another. It is worth emphasizing that the operational definition of complexity adopted above is not universal but does describe a large class of phenomena having *1/f*-variability, first observed in a physical context and named *1/f*-noise. The power spectrum of fluctuations is an IPL in frequency *f* of the form $S_p(f) \propto f^{-\beta}$ and has a long history of experiments and theories for $\beta = 1$, i.e. for *1/f*-noise, the vast majority of which is not relevant to the problems of interest here. We confine our remarks to biomedical phenomena that are rife with time series having IPL spectra with an index in the range $0.5 \leq \beta \leq 1.5$.

The concept of CS is at the apex of a hierarchical taxonomy[7] of the dynamic interactions of complex interacting organ networks. For this reason, we devote some time to developing a brief historical perspective on how that taxonomy unfolded. However, we emphasize the modifier 'brief' in that only the highlights of this historical record are noted here, which are focused on those properties most amenable to clarifying the information exchange among the brain, cardiac and respiratory organ networks.

**Normal synchronization**: The idea of synchronization, separate from that of complexity, was given its first articulation in a 1665 letter by the Dutch physicist Huygens. He observed that two pendulum clocks hung from the same beam but with independent initial conditions synchronize to one another in under thirty minutes. He referred to this anti-phase synchronous motion as "the sympathy of two clocks". Huygen's physical concept of synchronization predates Newton's Principia and therefore there was not even the vocabulary of mechanical forces with which to understand the phenomenon. Over the more than three centuries since Huygen's observation, his sympathy of motion has developed into what we understand today using a synthesis of the advances made in the understanding of nonlinear dynamics at its nexus with many-body dynamic networks.

At the turn of this century Strogatz chronicled in his excellent book[8] the evolution of the synchronization concept and from which we freely draw for the following few remarks. In the 1950's the mathematician Norbert Wiener[9] in his collaborations on the operation of the human brain identified the interaction of a spectrum of frequencies as being the basis of human consciousness. But it was the mathematical biologist Art Winfree[10] who in the 1960's identified the fundamental nature of nonlinear oscillators' interactions, which through critical dynamics produced transitions from disordered random behavior to highly ordered synchronous motion. In this way Winfree was able to identify dynamic self-organization as leading to biomedical synchronization as in circadian rhythm, the entrainment of pacemaker cells in the sinoatrial node of the beating heart, and elsewhere in the body's physiological organ networks. Winfree explained that the resulting synchronization produced an alignment in time as distinct from the spatial alignment previously observed in physical phase transitions, such as in the transition of a fluid to a solid.

In the 1970's Kuramoto[11] devised a simplified oscillator model of self-organization in time that included the related insights of both Wiener and Winfree, but with a symmetric interaction among the oscillator modes. The symmetry assumption enabled Kuramoto to obtain analytic solutions and thereby be the first to determine that a population of entities, from fireflies to brain cells, must have sufficiently similar properties to synchronize their complex dynamics. While the individual oscillators in the Kuramoto model are regular the emergence of global synchronization is independent of whether the individual oscillators are regular or stochastic.

The term normal synchronization refers to the entrainment of the dynamic equations of two or more interacting networks. Consequently, this would include the critical dynamics of many-body phase transitions.

**Synchronization and information exchange**: The Complexity Matching Hypothesis (CMH)[5] was a major step in understanding the nature of complexity in daily-life and states that the information exchange between interacting networks is optimal when the level of complexity of the two networks are the same[12]. This information flow has also been identified with 'strong anticipation' and in their nomenclature the dynamics have been interpreted as 'anticipation synchronization' resulting from the coupling between 'master' and 'slave' networks[13]. Marmelat and Delignieres[14] applied this interpretation to the interpersonal synchronization through a matching of the complexity indices (fractal dimensions). Along with colleagues they[15] also examined the CMH using a method based on the correlation between multifractal spectra, considering different ranges of time scales. They could distinguish between "situations underlain by genuine statistical matching, and situations where statistical matching results from local adjustments". They analyzed empirical datasets of biannual coordination, interpersonal coordination, and walking in synchrony with a fractal metronome. Note that the interacting networks need not be from the same scientific discipline in that network models are not discipline specific, and therefore the same model structure can be applied to networks of different origins across various disciplines (i.e., the last being a transdisciplinary phenomenon).

When two complex networks interact, information flows from the more complex (driver) to the less complex (driven) network. The driver perturbs the driven, and when the changing complexity of the driven becomes equal to that of the driver, as measured by the IPL indices $\mu_{driven}$ and $\mu_{driver}$ of the two networks, the maximal transfer of information occurs[13,16]. Mahmoodi et al.[17] established that scaling synchronization is a consequence of the fact that a very large number of crucial events for the IPL index in the interval $2 < \mu < 3$ using the modified diffusion entropy analysis (MDEA) data processing technique, which is reviewed in the Suppl. Info (SI.1). This notion of complexity matching has developed into the principle of complexity management (PCM)[18] to include the influence of one network on the other when the level of complexity of the two networks are very different, which brings us to the concept of CS.

The main difference between PCM and CS is that PCM rests on adapting the linear response theory to the case where the perturbed system is characterized by an IPL index[18], making it nonstationary, and the influence of perturbation does not affect the IPL of the single trajectories, but the average over infinitely many responses to the same perturbation. CS, on the contrary, is realized as a surprising synchronization between the scaling of single trajectories. This effect seems to be incompatible with the assumption that two networks interact without influencing their own complexities. In other words, networks can be initially out of scaling synchronization, and the transfer of information can directly change their IPL index. Experimentally, the change in complexity was observed in the rehabilitation of the elderly in the walking arm-in-arm experiment[16]. Although PCM was invoked to explain this rehabilitation effect, Mahmoodi et al.[17] made a model for this therapeutical synchronization, generating a surprising similarity between numerical and experimental results. The two interacting systems in this model change their complexity to realize the experimentally observed coordination. We believe that the CS illustrated in this paper goes behind the limits of PCM and is closely connected to the model of Mahmoodi et al.[17].

# Results

In this section we present a new way to characterize how the brain can exchange information with two other major physiological organ networks, as depicted in Figure 1, those being the respiratory and cardiovascular. In this figure are depicted the power law scaling indices for the processed time series, as discussed in the Methods section, from each of the 64 channels of a standard EEG, along with the those from ECG and respiratory organ networks that were simultaneously measured. It is clear from the figure that the quasiperiodic behavior of the scaling indices from the EEG channels are in synchrony with those from the ECG and respiratory organ networks.

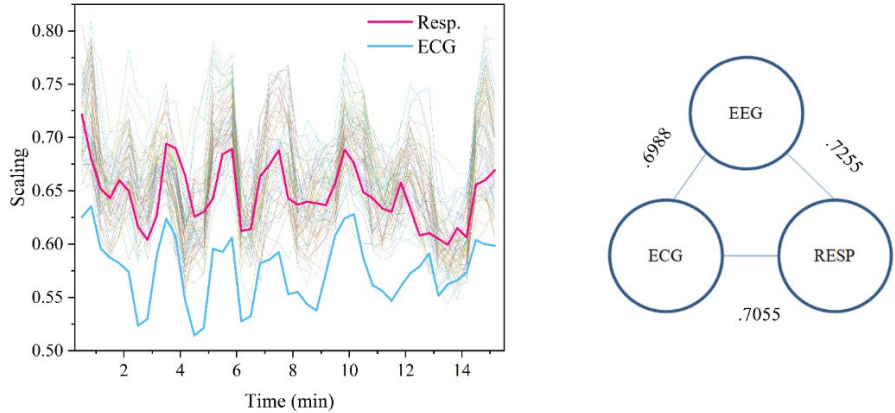

Figure 1. Left panel: Lighter curves are the scaling over time of 64 EEG channels. The red and blue curves are the scaling over time of the respiration and ECG organ networks, respectively. The individual datasets were processed using MDEA (see Methods and SI.1), on data windows of 1 minute duration (windows in increments of 20 s steps). The data was collected while the participant was engaged in a Neurofeedback task, as described in the Supplementary Information. Right panel: the corresponding cross-correlations between the average scaling over time of the EEGs and of the ECG and Respiration channel.

The time series $X_j(t)$ is the output from channel $j$ and scales when the time $t$ is multiplied by a constant $\lambda$ resulting in; $X_j(\lambda t) = \lambda^{\delta_j} X(t)$, and consequently, the time series is a homogeneous scaling function. Note that the scaling index $\delta_j$ is a local quantity in time as explained in (SI.1) and is related to the fractal dimension of the 'network' in the brain generating the time series in the vicinity of channel $j$. The time series underlying the scaling index in each of the EEG channels is seen in the figure to be multifractal with a quasiperiodic time dependence, with a similar interpretation for the cardiac and respiratory time series. The processing of the underlying time series is given in (SI.1). Figure 1 provides information regarding the way the dynamics of the entire brain, the heart and the lungs influence one another. The scaling indices of all 64 EEG channels are compared with the scaling index for the cardiovascular organ network (blue curve) and the scaling index of the respiratory organ network (red Curve). To properly interpret the behavior depicted in the figure requires that we answer the question: What is a scaling parameter and what does it entail about the underlying dynamic network?

West and Grigolini[5] review how the IPL indices for the PDF, given by $\psi(t) \propto t^{-\mu}$ and for the power spectral density (PSD) given by $S_p(f) \propto f^{-\beta}$ are related by $\beta = 3 - \mu$. The scaling index $\delta$ of the above homogeneous scaling relation determines that of the IPL index for an asymptotically ergodic time series by $\mu = 1 + 1/\delta$ as recorded in Table 1. Consequently, either $\beta$ or $\mu$ can be used as measures of complexity by means of the scaling index $\delta$. For ergodic time series, such as that asymptotically determined by the IPL index, $\mu$ increases with decreasing scaling index $\delta$ and complexity decreases along with $\delta$.

The scaling (complexity) of brain activity resulting from the empirical time series considered here is seen to be typically greater than respiratory scaling (complexity) which is typically greater than that of cardiac scaling (complexity). However, subsequent comparisons of the changes in scaling over time for the brain, heart and respiratory organ networks indicate that although the brain appears to have the greatest potential complexity the complexity-ordering with other physiologic organ networks depends on multiple factors, e.g. health status, cognitive states and task demands.

|  | Scaled functions | Parameter relations | Parameter range |  |
| --- | --- | --- | --- | --- |
| Waiting-time PDF | $\psi(t) \propto t^{-\mu}$ |  | $1 \leq \mu \leq 3$ |  |
| Power spectrum | $S(f) \propto f^{-\beta}$ | $\beta = 3 - \mu$ |  |  |
| Scaled variable | $X(t) \propto t^{\delta}$ | $\delta = \mu - 1$ | $1 \leq \mu \leq 2$ | non-ergodic |
|  |  | $\delta = 1/(\mu - 1)$ | $2 \leq \mu \leq 3$ | ergodic |
|  |  | $\delta = 0.5$ | $\mu \geq 3$ |  |

Table 1: This table makes easy reference to the scaling index δ from the above homogeneous scaling relation for the scaled variable *X(t)*; relates it to the PSD $S_p(f)$ index β through the waiting-time PDF $\psi(t)$ index μ in the two asymptotic regimes. The value μ=2 is the boundary between the underlying process having a finite (μ>2) or an infinite (μ<2) average waiting time and is also the point at which *β=1* where the process is that of *1/f*-noise. An excellent review of the close relationship between complexity of a network and the presence of fractal fluctuations (*1/f*-variability) in its macroscopic behavior is given by Delignieres and Marmelat[19].

The correlations noted here do not pertain to the underlying time series generated by the organ networks, but to the scaling indices of those time series thereby being part of the new area of investigation alluded to earlier. Figure 2 shows the maximum of the cross-correlation function between the time series for the scaling indices of the 64 EEG channels, the respiratory and cardiac organ networks depicted in Figure 1. It is evident from Figure 2 that the maximum cross correlation between the EEG channels do not decay monotonically in time as the separation distance increases between the channel-probes attached to the scalp. This result suggests that channels on different sides of the brain may be more strongly coupled than those that are side-by-side depending on the relative functions of the brain regions. Buzsaki[20] suggests that a "small-world-like" strategy may be operative in the architecture supporting the intra-brain information transfer, see Strogatz[9] for an excellent review of small-world theory in the present context. Buzsaki goes on to say that both theory and modeling suggest that long-range interneurons are critical for brain-wide synchronization of gamma and potentially other oscillations.

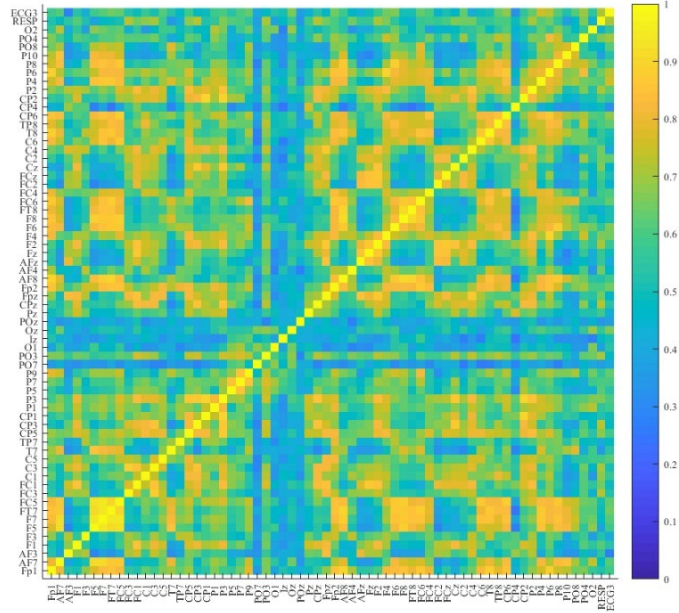

Figure 2: The maximum cross correlation between the 64 + 2 (64 EEG channels, ECG, and Respiration) scaling time series depicted in in Figure1. Note that this cross-correlation matrix can also be interpreted as 66-node dynamic network and network theory can be applied to the interpretation of the three organ mutual dynamics.

Recall that the scaling index $\delta$ and complexity rise and fall together, but not in direct proportion to one another, since a phenomenon and its measure are not the same thing. One may therefore interpret Figure 2 as a discrete map of the changing complexity within the brain, and potentially provides a new measure of the flow of information, see Discussion section. What is evident from this figure is that identifying distant channels gives information about the level of synchronization of brain regions from which we ought to be able to infer the existence of the connectivity of functionality. This suggests that we can talk about a new measure of functional connectivity, but it is perhaps premature to discuss the flow of information without having multiple time-dependent complexity maps from which to determine the changing functionality role of a given channel.

In the right panel of Figure 1 is recorded the maximum of cross-correlations between the instantaneous average of the scaling index over all 64 EEG channels (to obtain an average scaling index for the brain), with the scaling indices for the cardiovascular organ network (blue curve in left panel) and the respiratory organ network (red curve in left panel). The maxima of the cross-correlation functions calculated are approximately the same for all three combinations. This result suggests that in a healthy individual the complexity of the organ networks is approximately the same on average, so that information flow may well satisfy the CMH on average.

## Discussion

We have defined complexity in terms of the IPL index for the waiting-time PDF for CEs, a form of self- organized temporal complexity[21,22] and shown elsewhere[17] that complexity change results from the interaction between at least two complex networks. We have processed CE time series and determined that the IPL index is time dependent for interactions between physiologic organ networks. This is distinct from the forms of coupling identified for the interaction of the same organ networks studied by Bartsch and Ivanov[6], during different stages of sleep. They[6] determined that the interaction between the brain, cardiac and respiratory organ networks reveal pronounced phase transitions associated with detailed reorganization in network structure and links strength in response to changes in the underlying neuroautonomic regulation. Further, they identified the multifractal nature of the time series generated by the physiologic organs and identified new and interesting forms of coupling using linear coherence measures. However, lacking the data processing technique provided by MDEA, or its equivalent, they did not uncover the high-order synchrony of the time dependence of the scaling indices of the brain, cardiac and respiratory organ networks.

Delignieres and Marmelat[19] suggested that a network's complexity "and the presence of *1/f* -fluctuations in its macroscopic behavior have opened new domains of investigation". One such domain is entailed by the multifractal nature of the scaling indices depicted in Figure 1. The multifractality indicates that all three (different parts of the brain, heart, and lungs) organ networks have dramatic changes in complexity over time, as indicated by the quasi-periodic time dependence of the scaling index, being a direct consequence of their inter-network and intra-network interactions. There is no set hierarchy among the scaling indices for parts of the brain and the physiologic organ networks they monitor and with which they exchange information, see SI.3. The brain receives information from various sensor organs. For example, on 9/11 people saw smoke billowing from the two towers; as the towers collapsed, they smelled the heat and choked on the clouds of dust and debris; they heard the screams of those not realizing what was happening to them. The brain processes the sensor signals and from the synthesis of the processed information, makes an executive decision of either 'flight or fight', ultimately transmitting the decision for action to the motor control network. The average person fled from the site, while the first responders fought their way into the devastation. The drama of that day highlights not only that the human brain receives information from the five senses of sight, sound, smell, taste and touch, but subsequently transmits information to motor control networks in response to its processing of the information received. In the final analysis the brain determines whether you make it to safety or have your name inscribed on a granite slab.

Note that the scaling indices are changing over time in Figure 1, thereby indicating the multifractal nature of the time series from the brain, cardiac and respiratory organ networks. This characterization of the dynamics suggests a parallel with Huygens' observations of the "sympathy of two clocks", in that his observation lacked the theoretical foundation subsequently provided by Newton's mechanics, and we also lack a comprehensive theory with which to explain the necessity for this manner of controlling physiologic signals. However, unlike Huygens we have the advantage of a three-century perspective of how to use the scientific method to develop theory necessary to further understanding. In this regard we call attention to the neuroscientist Buzsaki[23], who observed that transient coupling between various parts of the brain may support information transfer as described by 'small world theory' and the scaling results shown in the figure support this conjecture. But a word of caution is appropriate here because the synchrony observed in the scaling indices is not directly related to the synchronous behavior observed from the correlation properties of the time series.

We established the connection between multifractal and CE in surrogate time series[24]. Elsewhere[25], we found the same connection in physiologic time series using heartbeat data, thereby supporting the fundamental role that CEs play in the exchange of information between interacting complex organ networks. Therein we established the consistency of two apparently different diagnostic techniques. The first technique is based on the multifractal spectrum of healthy individuals being broader than that of pathologic subjects. The second technique is based on heartbeat dynamics being a mixture of CEs and non-CEs, with pathologic patients having a high probability of hosting a greater percentage of non-CEs. Moreover, we proved therein that increasing the fraction of non-CEs in the heartbeat time series reduces the width of the multifractal spectrum thereby simultaneously establishing compatibility of the two techniques, while providing a dynamic interpretation for the source of multifractality as being due to the level of complexity of the organ network, using the $1/f$-variability scaling index as the complexity measure.

We also need to point out that there is a distinct difference between chaos synchronization and complexity synchronization (CS). A single nonlinear dynamic network in the process of chaos synchronization is chaotic and surprisingly two such networks can synchronize while simultaneously maintaining the chaotic dynamics they had in isolation. Chaos synchronization is separate and distinct from CS so we spend a little time recalling the main features of the former in SI.3 to avoid confusion in discussions of the latter.

The complexity map depicted in Figure 2 suggests that it would be of value to identify EEG channels that are physically distant from one another and yet are freely exchanging information, perhaps in the form of a 64-node information network with links of varying strength over time. CS is observed in a variety of diverse tasks such as neurofeedback and shooting (see SI.2). We emphasize that the MDEA method could show the interactions between complex networks which are difficult to analyze with traditional methods due to non-stationarity and/or different time scales of used measures, i.e., ECG, EEG, and respiration. Moreover, it can potentially be a very valuable approach in investigating brain-body interactions in other hard to analyze networks, e.g. the brain-gut[26].

To observe the different complexities in the organ networks analyzed herein we may need to design new experiments with some learning/training phase (not investigated in this study). This would enable us to observe which network is the driver and which is the driven in the manner used so successfully by Ref.[6].

## Methods

**Participants:** Thirty (N = 30; 13 female) young healthy adults (ages 18-40 yr; mean 24.99 ±3.21) participated in five separate sessions within a three-week interval. Volunteers who agreed to participate were asked to read and sign an Informed Consent Agreement (approved by the Human Use Committee at the US Army Research Laboratory and the Institutional Review Board at the University of Maryland, Baltimore County, in accordance with the Declaration of Helsinki and the U.S. Code of Federal Regulations). For this proof-of-concept paper, we selected one participant from one session whose post-processed dataset was artifact-free.

**Neurofeedback Training:** Neurofeedback (NF) training was implemented using an HTC Vive virtual reality system (https://www.vive.com/us/). The frontal midline theta (Fmθ) feedback signal was derived from channel FCz. FCz was referenced to averaged mastoids (A1/A2), bandpass filtered (0.5-30 Hz), and power spectral density was estimated over 1-s windows (2048 samples) with 875 ms overlap (1792 samples) and averaged over 4-7 Hz (8 estimates/s with 1 s history was visually displayed during neuromodulation). Each session consisted of six blocks of six trials, and each trial consisted of a 10-s rest period followed by a 30-s active modulation period. Two NF training groups were tested for differential effects on executive task performance: an increase Fmθ group (INC) and an alternating increase/decrease Fmθ group (ALT). Participants in the INC group (n=12) were instructed to increase Fmθ every trial, block, and session, whereas those in the ALT group (n=18) were instructed to alternate between increasing and decreasing Fmθ across blocks of trials (randomly determined) each session (for more details see Kerick et al.[27], Spangler et al.[28]).

**Signal Acquisition and Processing:** All data were acquired simultaneously at 2048 Hz and referenced online to the Common Mode Sense (CMS) and Direct Right Leg (DRL) electrodes using a 64 (+8 external) channel BioSemi system (Amsterdam, The Netherlands; http://www.biosemi.com/products.htm). An integrated auxiliary channel was used to acquire respiration data (Nihon Kohden TR-753T respiration belt). Signal processing of electroencephalographic (EEG), electrocardiographic (ECG), and respiration (RESP) time series was applied using EEGLAB (ver 14.1.2b; http://sccn.ucsd.edu/eeglab/) and in-house code using MATLAB (9.3.0.713579; Natick, MA) on a 64-bit Linux operating system. EEG and ECG data were high pass filtered at 1 Hz and RESP data were lowpass filtered at 2 Hz using zero-phase finite impulse response filters. EEG data were re- referenced to the average of all 64 EEG channels, down sampled to 512 Hz, and cleaned of artifacts using independent component analysis[29]. ECG and RESP data were also down sampled to 512 Hz after filtering.

**Theory:** We can now answer the question posed earlier: What is a scaling parameter and what does it entail about the underlying dynamic network?

For a stochastic process the dynamic variable *X(t)* with homogeneous scaling is given by $X(\lambda t) = \lambda^\delta X(t)$ and which is interpreted in terms of the scaling PDF:

$$P(x,t) = \frac{1}{t^\delta} F\left(\frac{x}{t^\delta}\right), \qquad (1)$$

where $P(x,t)dx$ is the probability that the random variable *X(t)* is in the interval (*x, x + dx*) at time *t*. The PDF function *F(.)* is unknown in general, however for *δ = 0.5* the PDF is Gaussian, and the process is diffusive. If the PDF is Gaussian but δ is not equal to 0.5 the process is said to describe fractional Brownian motion (FBM), as first described using the fractional calculus by Mandelbrot and van Ness[1]. The more interesting case is when the unknown function is not Gaussian, for example, when the process is Lévy stable. Using the definition of Wiener/Shannon entropy we obtain using the scaling PDF, without knowing the *F(.)* function, the deviation of the entropy from its reference state defined by the unknown function is:

$$\Delta S(t) = S(t) - S_{ref} = \delta \ln(t). \qquad (2)$$

Consequently, a graph of the entropy for such a process versus the logarithm of the time is a straight line whose positive slope gives the scaling index. The details of the MDEA processing method applied to the three datasets to construct diffusive trajectories and the corresponding entropy are outlined in SI.1. The time derivative of the entropy (information) given by Eq. (2) asymptotically vanishes as an IPL with unit index, entailing that the average of the stochastic variable asymptotically becomes constant.

Diffusion entropy analysis (DEA) was originally introduced to study the complexity of a social process[30-32]. There is a significant difference between the DEA and its generalization to modified DEA (MDEA) from other entropy. The DEA records a positive unit step in the diffusion trajectory for every event in the dataset whether it is a CE or not. Many of these diffusion trajectories for a given dataset are collected into an ensemble, the ensemble PDF is calculated from the histogram and subsequently inserted into the expression for the Shannon/Wiener entropy. The resulting time-dependent entropy is plotted against the log of time and yields the scaling index δ as predicted by Eq. (2). The fact that the empirical time series contains both CEs and non-CEs is a limitation on the technique which is resolved by the generalization to MDEA. The modification is to partition the time axis into many equal- sized time intervals (stripes) and record each crossing of one stripe to the next to construct the new diffusive trajectory. This use of stripes has the effect of filtering out the influence of the non-CEs in the empirical time series on the diffusion trajectory, resulting in a more accurate measure of the scaling index over a longer time than obtained using DEA.

## Conclusions

A major result of the present work is that we have shown that when the scaling parameter δ is greater than 0.5 there is a fork in the road beyond δ = 0.5 leading to two distinctly different understandings of the dynamics of the complex network being considered. The time series generated by an organ network beyond the fork would have two very different kinds of statistics depending on which branch is found from the data. The branch of one leg beyond the fork is interpreted by Mandelbrot to be fractional Brownian motion (FBM)[33], which has Gaussian statistics and an infinite memory. Whereas the other branch of the fork has the statistics interpreted by the Grigolini-West group[5] to be crucial events (CEs), in which the statistical fluctuations are renewal, and the PDF is given by the scaling form of Eq. (1) with a quite general non-Gaussian form. It is this second branch that describes the dynamics of healthy complex organ networks. It is not just that other investigators assume that organ networks have statistical properties described by FBM[6,14-16,19,33], which they do, but that none of them to the best of our knowledge, even tip their hat to the possible existence of the CE branch.

The results of the present paper establish that broad multifractal spectra transfer information between complex organ networks by means of CEs. This is consistent with earlier findings of our group regarding the connection between the multifractal spectrum and SOTC fluctuations[17] and subsequent findings[5,24,26] all supporting the surprising conclusion reached in the above paragraph. Consequently, this alternative interpretation of the statistics fulfills the promise made by Ivanov et al.[34] of the new fields of network physiology and network medicine giving rise to significant progress in understanding biomedical phenomena.

In summary we list the theoretical findings supported by the diffusion entropy analysis (MDEA) of the time series generated by the interacting brain, heart and lung organ networks:

1. The complexity level of time series generated by organ networks is measured by the scaling index $\delta$.
2. Time series generated by organ networks *X(t)* are homogeneous random fractals having a self-similar PDF given by Eq. (1).
3. The scaling index is typically time-dependent $\delta(t)$ making the organ-network time series multifractal.
4. Complexity synchronization (CS) is the mechanism necessary to coordinate the time-dependent scaling behavior between interacting organ networks.
5. It is worth speculating that the techniques developed in Network Science may lead to a significantly deeper understanding of connective functionality of the brain than has been provided by less robust modeling techniques.

Although consistent with CS, we cannot say definitively from the data presented here (two subjects performing two different tasks) that one network is driving another. Future research is needed to further investigate factors which may influence the hierarchical/driver/driven.

## Acknowledgements
This study was supported by the US Army Research Laboratory cooperative agreement W911NF-16-2-0008.

## Author contributions
S.E.K. conceived and conducted the original experiment, K.M. conceived CS, B.J.W wrote the paper. All authors contributed to analyses of results and editing and reviewing the manuscript.

## declaration of competing interests
The authors declare no competing interests.

## Data and Code Availability
All data and code are available upon request (K.M. or S.E.K.)

# Supplementary Information

**SI.1**: **Modified Diffusion Entropy Analysis (MDEA):**

MDEA was applied to post-processed continuous data from all 64 EEG channels, the ECG, and the RESP network of one participant in one session of neurofeedback NF training. Estimates of scaling indices were obtained on 60 s data windows in increments of 20 s steps, yielding a time series of scaling exponents for each channel. For MDEA each of the datasets was projected onto the interval [0,1] which was then divided into parallel stripes of size 0.01 (panel a of Figure SI.1-1, ECG data.) Next, the events were extracted, and defined as 1 if the signal at that time is in different stripe with respect to its previous value (panel b of Figure SI.1-1). Using the time series of the extracted events, we created a diffusion trajectory (panel c of Figure SI.1-1 i.e., the cumulative sum of the events in panel b).

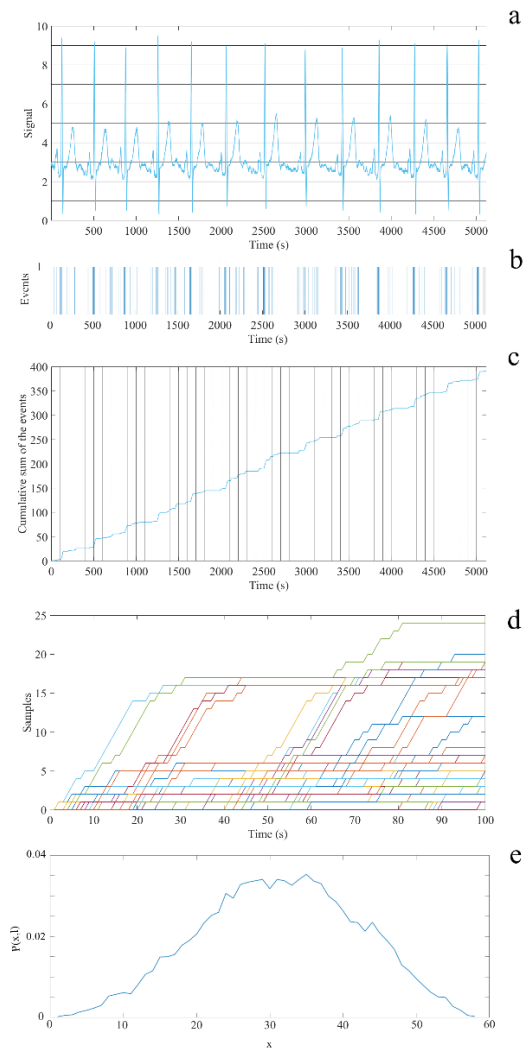

Figure SI.1-1. The schematic of the Modified Diffusion Entropy Analysis (MDEA). Panel a) the blue curve is the heart rate signal which was first projected to the interval [0,1] and then divided into stripes of size of 0.01. The horizontal lines define the stripes. Panel b) The events (represented with ones), were extracted from the passings of the blue curve from one stripe to others. Panel c) The diffusion trajectory made by cumulative summation of the events of panel c. The vertical lines show a selected set of windows with sizes 100 that sliced the diffusion trajectory. Panel d) The sliced trajectories of panel c were shifted to start from the origin. Panel e) The histogram of the position of the trajectories at the end of the windows (to create this histogram we used 60 sec of data and stripe size of 0.01, which are the DEA parameters used.)

To make statistics of a single diffusion trajectory, we pick a window size *l* and slice the signal into many pieces, each length *l* and starting from an event (panel d). By shifting all the slices to start from origin, we can evaluate the distribution of trajectories at time *l* (panel e). Using these distributions for different window sizes we can define Wiener/Shannon Entropy, assuming that $P(x, l)$ is the PDF corresponding to window size *l*, to be:

$$S(l) = - \int P(x,l) \ln P(x,l) dx. \tag{3}$$

If the PDF is given by Eq. (1) with *t* replaced with *l* we obtain Eq. (2) with the same replacement. So, the slope of $\Delta S(l)$ vs. $\ln l$ will give the scaling index δ, as shown in Figure SI.1-2 for three different signals.

After evaluating the scaling time series for 64 EEG channels, ECG and Respiratory network data, using MDEA, we studied the cross-correlations between them, see Extended Data. The results confirm the strong connection between the complexity of the three systems.

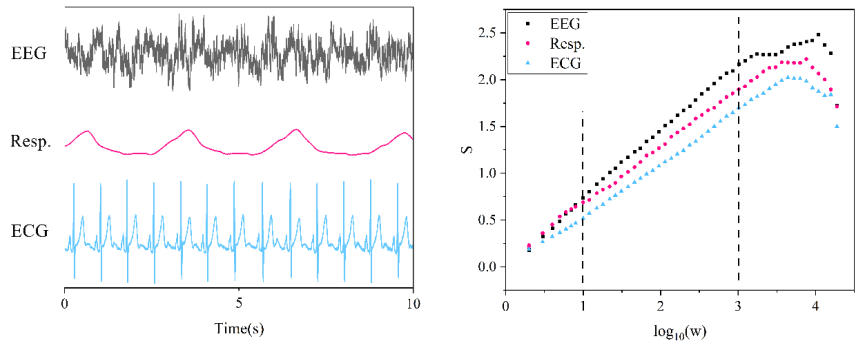

Figure SI.1-2: Left panel: 10 seconds of EEG, respiration, and ECG data. Right panel: The corresponding MDEA graphs of the time series on the left panel. The slopes of the linear parts are the measure of temporal complexity of the time series.

**SI.2 Extended Data (Go-NoGo shooting task)**

The Go-NoGo task was implemented in virtual reality using HTC Vive (https://www.vive.com/us/). In each session, the participants completed four blocks of 90 trials (360 total trials) in each low and high time-stress condition (2160 total trials in each condition over six sessions). Pop-up targets were pseudo randomly distributed 40 times at each of 9 range locations (three simulated distances (near, mid, far) by three lanes (left, center, right) and exposed at variable onset intervals (1000 ± 500 ms over a Gaussian-distributed range of 100 ms increments) for various target exposure durations. The probability of targets (enemy; red) to non-targets (friendly; green) was .90/.10, respectively to induce a prepotent response bias.

The participants were instructed to "shoot enemy targets as quickly and accurately as possible, while refraining from shooting at friendly targets". Time-stress conditions were individualized based on a pre-testing performance thresholding procedure to account for individual differences in participants' ability to perform the task. This was done by empirically determining target exposure durations corresponding to the 50th (high time-stress) and 90th (low time-stress) percentile hit-rates in response to 100 enemy targets using psychophysical methods (method of limits). Herein we analyze scaling indices of EEG, ECG, and RESP time series data for the same participant during the Go-NoGo shooting task as was applied for data recorded during neurofeedback training (See figures SI.2-1 and SI.2-2).

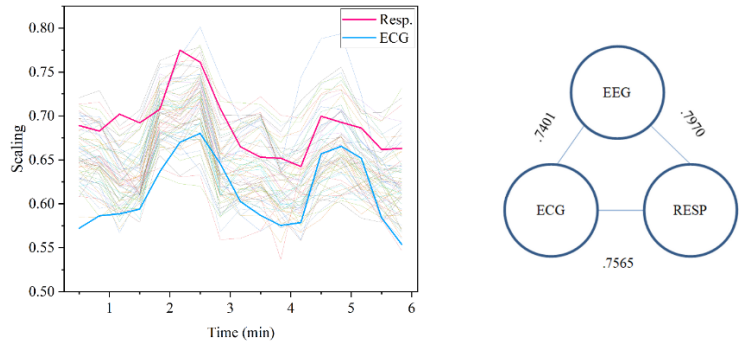

Figure SI.2-1: Left panel: the scaling over time obtained by MDEA for 64 EEG channels (lighter curves), respiration (red curve), and ECG channel (blue curve). Right panel: the corresponding cross-correlations between the average scaling over time of the EEGs and of the ECG and Respiration channel.

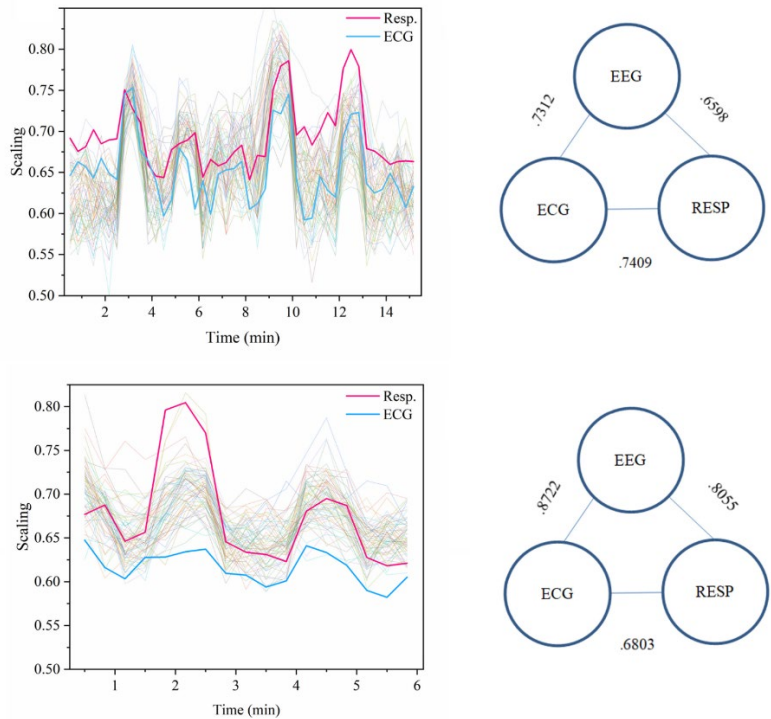

Figure SI.2-2: Top-left and bottom-left panels are the scaling over time (obtained by MDEA) for another participant during Neurofeedback task and while doing Shooting task, respectively. The lighter curves represent the EEG channels, and the red and blue curves represent the respiration and EEG channel, respectively. The right panels show the corresponding cross-correlations between the average scaling over time of the 64 EEGs and the ECG and Respiration channel.

**SI.3 Chaos Synchronization**

Chaos appears in most nonlinear dynamic systems, whether in discrete mappings, in Hamiltonian systems, or in dissipative systems. An essential feature of chaos is its sensitive dependence on initial conditions such that the final state has a finite predictability horizon, which can be interpreted to mean that two solutions (trajectories) to the deterministic equations of motion with arbitrarily small separation of initial conditions exponentially separate from one another and the final sates of the two solutions are arbitrarily distant from each other.

A solution is pulled from an arbitrary initial condition to an attractor, that being a manifold in a finite dimensional phase space on which the trajectory of the motion unfolds. A chaotic system, such as that of Lorenz, unfolds on a 'strange attractor'. The Lorenz attractor is embedded in a three-dimensional phase space and due to dissipation in one of its variables the motion along that dimension dies out asymptotically, and the trajectory is drawn to an attractor with a non-integer dimension greater than two but less than three. The result is a new kind of order – one in which the motion of the trajectory confined to the strange attractor is unpredictable[SI1]. It is the fractal dimension of the strange attractor that enables the trajectory to unfold on the fractal manifold in a non-periodic fashion.

It was thought for a long time that chaos was incompatible with synchronization, that is until Pecora and Carroll[SI2] decided to apply chaos theory to encrypting messages in a chaotic signal for communications. The chaotic fluctuations mask the message from the sender, which is retrievable by the receiver using the deterministic dynamics of a second chaos generator identical to the first. This strategy of driving a computer simulation of a receiver (a network with a strange attractor solution) with a chaotic signal transmitted from a duplicate of itself, was indeed sufficient to coax the two trajectories into synchronization. Subsequently, it was recognized[SI3] that synchronization of chaotic time series had been first accomplished in 1983 and by half-a-dozen other investigations prior to Ref.[SI4].

**SI References**